\documentclass{emulateapj}

\shorttitle{On the source of Faraday rotation in 3C~120}
\shortauthors{G\'omez et al.}

\begin{document}

\title{On the source of Faraday rotation in the jet of the radio galaxy 3C~120}

\author{Jos\'e L. G\'omez\altaffilmark{1}, Mar Roca-Sogorb\altaffilmark{1}, Iv\'an Agudo\altaffilmark{2,1}, Alan P. Marscher\altaffilmark{2} and Svetlana G. Jorstad\altaffilmark{2,3}}

\altaffiltext{1}{Instituto de Astrof\'{\i}sica de Andaluc\'{\i}a, CSIC, Glorieta de la Astronom\'{\i}a s/n, 1808 Granada, Spain. jlgomez@iaa.es; mroca@iaa.es}

\altaffiltext{2}{Institute for Astrophysical Research, Boston University, 725 Commonwealth Avenue, Boston, MA 02215, USA. iagudo@bu.edu; marscher@bu.edu; jorstad@bu.edu}

\altaffiltext{3}{Astronomical Institute, St. Petersburg State University, Universitetskij Pr. 28, Petrodvorets, 198504 St. Petersburg, Russia}

\begin{abstract}

  The source of Faraday rotation in the jet of the radio galaxy 3C~120 is analyzed through Very Long Baseline Array observations carried out between 1999 and 2007 at 86, 43, 22, 15, 12, 8, 5, 2, and 1.7 GHz. Comparison of observations from 1999 to 2001 reveals uncorrelated changes in the linear polarization of the underlying jet emission and the Faraday rotation screen: while the rotation measure (RM) remains constant between approximately 2 and 5 mas from the core, the RM-corrected electric vector position angles (EVPAs) of two superluminal components are rotated by almost $90^{\circ}$ when compared to other components moving through similar jet locations. On the other hand, the innermost 2 mas experiences a significant change in RM -- including a sign reversal -- but without variations in the RM-corrected EVPAs. Similarly, observations in 2007 reveal a double sign reversal in RM along the jet, while the RM-corrected EVPAs remain perpendicular to the jet axis. Although the observed coherent structure and gradient of the RM along the jet supports the idea that the Faraday rotation is produced by a sheath of thermal electrons that surrounds the emitting jet, the uncorrelated changes in the RM and RM-corrected EVPAs indicate that the emitting jet and the source of Faraday rotation are not closely connected physically and have different configurations for the magnetic field and/or kinematical properties. Furthermore, the existence of a region of enhanced RM whose properties remain constant over three years requires a localized source of Faraday rotation, favoring a model in which a significant fraction of the RM originates in foreground clouds.
  
\end{abstract}


\section{Introduction}

  Polarimetric Very Long Baseline Interferometric (VLBI) observations have proven to be a very powerful tool to probe the magnetic field structure in relativistic jets of Active Galactic Nuclei (AGN). However, recent observations have revealed that the actual orientation of the polarization angle at parsec scales may be severely affected by Faraday rotation \citep[e.g.,][]{Udomprasert:1997p14601,Taylor:1998p14529,Reynolds:2001p2238,Asada:2002p1398,Zavala:2003p2289,Zavala:2004p2285,Attridge:2005p1401,Gomez:2008p1527,OSullivan:2009p8966,Kharb:2009p9578,Asada:2010p13535,Taylor:2010p13627}, which rotates the polarization vector by an angle proportional to the square of the observing wavelength. The analysis of Faraday rotation in jets of AGN is therefore of special relevance to properly determine the actual orientation of the polarization angle, and hence of the magnetic field structure on the plane of the sky in the emitting region.
    
  Several possibilities have been suggested as the source of the observed Faraday rotation measure (RM) in jets of AGN. Gradients in the RM along the jet \citep[e.g.,][]{Taylor:1998p14529,Gabuzda:2003p1456,Gomez:2008p1527} and variability in both space and time \citep[e.g.,][]{Zavala:2001p2284,Zavala:2003p2289,Zavala:2005p2288,Asada:2008p1399,Mahmud:2009p11368} suggest that the observed RMs are intrinsic to the inner few hundred parsecs of AGN. This rules out the possibility that the RM is produced by our own Galaxy, or in the host galaxy interstellar or intracluster medium. Internal Faraday rotation in the emitting jet would produce severe depolarization \citep{Burn:1966p1425,Cioffi:1980p1435} that is not observed, and can therefore be safely excluded as the cause of RM in most cases \citep[see however][]{Gomez:2008p1527,Homan:2009p9905}.
  
  The broad emission-line region (BLR) is thought to be smaller than a parsec and to have a small volume covering factor, so it can therefore be eliminated as the source of the RM that appears on scales of tens or hundreds of parsecs. The narrow emission-line region (NLR) extends to hundreds of parsecs, and could therefore in principle contribute to the observe RM, although its relevance is subject to uncertainties owing to its largely unknown volume covering factor \citep{Netzer:1993p14646}. The observed RM time variability has also been used as an argument to exclude the NLR, or the intercloud gas, as the main source for the RM \citep{Asada:2008p1399,Kharb:2009p9578}.
  
  First suggestion that the RM is produced in a sheath that is in close proximity with the jet was given by \cite{Asada:2002p1398}, based on the detection of a RM gradient across the jet width in 3C~273. Such transverse gradients are expected if the sheath is threaded by a helical magnetic field, since they can be produced by a systematic change in the net line-of-sight magnetic field component across the jet \citep{Laing:1981p1970,Blandford:1993p5382}. Observational confirmation for the existence of these helical magnetic fields is of special relevance, since it is thought that these may appear naturally through the rotation of the accretion disk or black-hole ergosphere from which jets are launched, and could have an important role in the actual formation, collimation, and acceleration processes \citep[e.g.,][]{Koide:2002p1957,Vlahakis:2004p10490,DeVilliers:2005p14727,Hawley:2006p14728,McKinney:2006p2147,Komissarov:2007p1960,McKinney:2009p9204,Marscher:2008p2014,Marscher:2010p12884,Jorstad:2010p13309}.
    
  Further observations have confirmed the transverse RM gradient in 3C~273, revealing also time variability on a scale of years \citep{Zavala:2005p2288,Asada:2008p1399}. Following this initial detection in 3C~273, other sources have also been claimed to contain transverse gradients in the RM \citep{Gabuzda:2004p1462,Gomez:2008p1527,Asada:2008p6013,OSullivan:2009p8966,Kharb:2009p9578,Croke:2009p12834,Asada:2010p13535}. \cite{Contopoulos:2009p10586} present a study of the observed RM in jets of AGN, claiming that 29 out of 36 sources possess transverse gradients in RM, further suggesting that the gradients have a preferred sense of direction. More recently, \cite{Taylor:2010p13627} have raised concerns about the claimed transverse RM gradients by arguing that (besides other reasoning based on jet opacity and significance of the measurements) most of these observations lack the necessary resolution transverse to the jet. \cite{Taylor:2010p13627} propose several criteria for establishing a reliable transverse RM gradient, claiming that only observations of 3C~273 have met these criteria so far. 
  
  The existence of a sheath that surrounds a fast, synchrotron-emitting jet is also supported by multiple numerical simulations \citep{Aloy:1999p1395,Aloy:2000p1396,Tsinganos:2002p2270,DeVilliers:2005p14727,Gracia:2005p5713,Gracia:2009p12333,Hawley:2006p14728,McKinney:2006p2147}. Furthermore, computation of the expected rotation measure from 3D general relativistic magnetohydrodynamic simulations by \cite{Broderick:2010p14820} show that it is possible to reproduce many of the RM structures in jets of AGN, including the presence of transverse RM gradients. There is, therefore, observational and theoretical support for the existence of a sheath that surrounds the emitting jet that may contribute significantly to the observed RM in jets of AGN. 
  
  Additionally, there is also strong observational evidence \citep[e.g.,][]{Oosterloo:2000p2227,SolorzanoInarrea:2001p2252,Middelberg:2007p12544} and theoretical support \citep[e.g.,][]{Steffen:1997p2257,Steffen:1997p2255,Wang:2000p2281,Bicknell:2003p14596,Saxton:2005p2247} for the existence of foreground clouds, probably interacting or entrained by the jet, that may contribute significantly to the observed RM \citep{Gomez:2008p1527,Mantovani:2010p13551}. In particular, \cite{Gomez:2008p1527} present a study of the radio galaxy 3C~120 consisting of 12 monthly polarimetric 15, 22, and 43 GHz Very Long Baseline Array (VLBA) observations revealing the existence of a localized region of high ($\sim6000$ rad m$^{-2}$) Faraday rotation measure between approximately 3 and 4 mas from the core. The detection of this region of high RM required the use of a novel method, consisting of the combination of the RM information gathered at all epochs, so that multiple superluminal components can be used to map the RM screen. This technique revealed a remarkably consistent RM across epochs. A smooth sheath around the jet in 3C~120 cannot produce a localized region of enhanced RM -- although it successfully explains the observed gradients in RM and degree of polarization along and across the jet -- so the RM was assumed to originate in a foreground cloud, presumably interacting with the jet.
    
  In this paper we present further observations of the radio galaxy 3C~120 aimed to obtain a better understanding of the origin of the RM screen in this source. Thanks to its proximity ($z$=0.033), VLBI observations are able to resolve the jet across its width, revealing a very rich and dynamic structure, even when observed at the highest frequencies \citep[e.g.,][]{Gomez:1998p1489,Gomez:1999p1485,Gomez:2000p1484,Gomez:2001p1526,Walker:1987p2272,Walker:2001p2273,Jorstad:2005p1946,Marscher:2002p2016,Marscher:2007p2015,Chatterjee:2009p11330,RocaSogorb:2010p13195}. 3C~120 is located at a distance of 140 Mpc ($H_0=71$ km s$^{-1}$ Mpc$^{-1}$), so that 1 mas corresponds to 0.64 pc. Superluminal motion of components along the jet provide an upper limit of the jet viewing angle of $\sim20^{\circ}$ \citep{Gomez:2000p1484}, so that distances along the jet appear appreciably foreshortened by projection.
  
\section{Observations and images}
\subsection{Observations and data reduction}

  We present VLBA observations of the jet in the radio galaxy 3C~120 corresponding to two different observing programs carried out in 2001 and 2007. Partial analysis of these data revealed gradients in Faraday rotation and polarization \citep{Gomez:2008p1527}, and unusually high brightness temperature 140 pc from the core \citep{RocaSogorb:2010p13195}.

\begin{figure*}
\epsscale{1.06}
\plotone{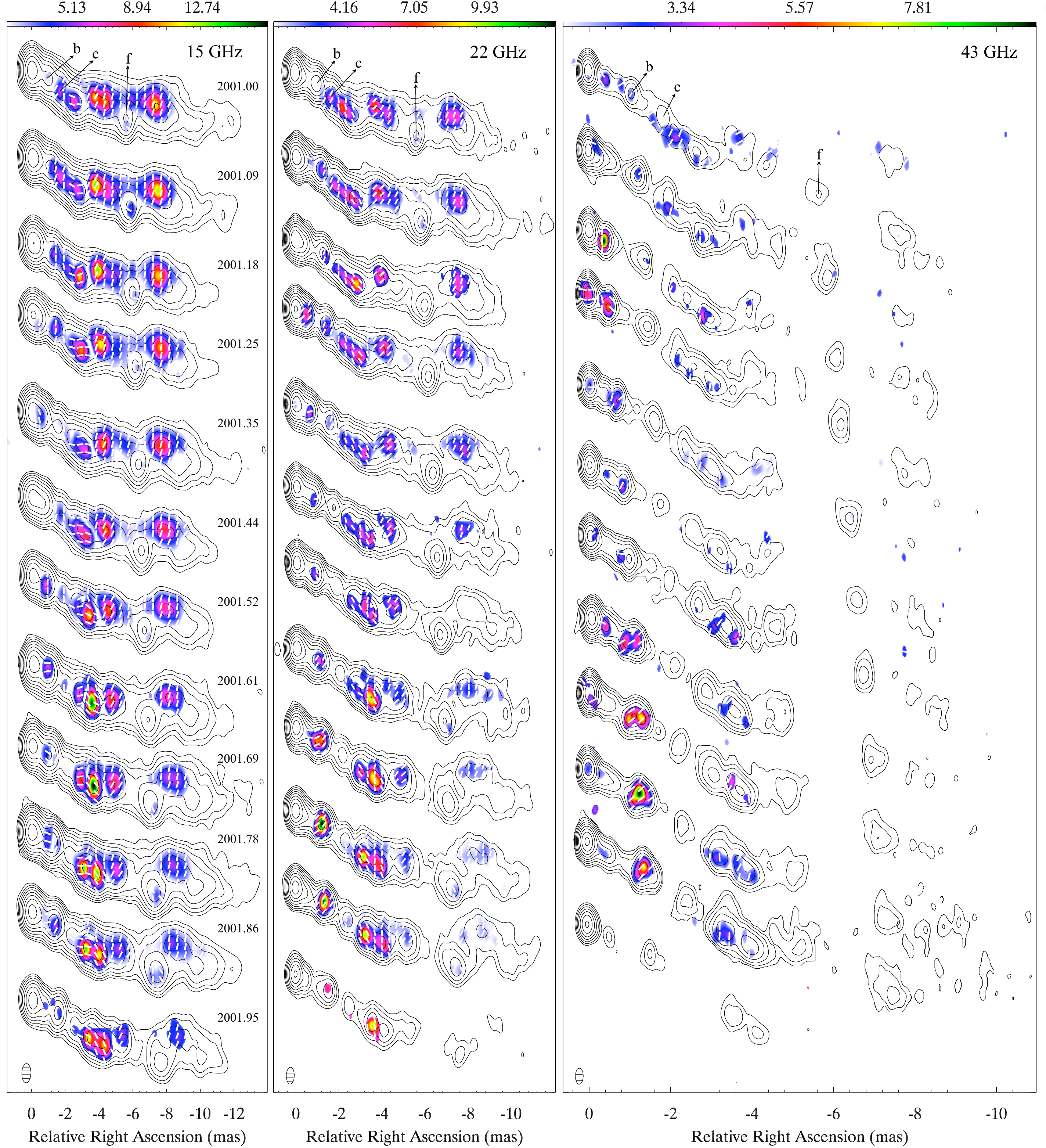}
\caption{15 (\emph{left}), 22 (\emph{middle}), and 43 GHz (\emph{right}) natural weighted VLBA images of 3C~120 in 2001. Note the different scale size used for each frequency. Vertical map separations are proportional to the time difference between epochs, shown in 15 GHz panel. Total intensity contours are overlaid at 0.5 (0.8; 1.0), 1.1 (1.7; 2.2), 2.5 (3.6; 4.6), 5.6 (7.7; 9.9), 13 (16; 21), 28 (35; 45), 62 (74; 95), 138 (158; 203), 310 (337; 433), and 692 (716; 923) mJy beam$^{-1}$ at 15 GHz (22; 43 GHz). A common convolving beam of FWHM $1.14\times0.54$ ($0.78\times0.37$; $0.40\times0.19$) mas at -1.3$^{\circ}$ (-0.9$^{\circ}$; -1.1$^{\circ}$) at 15 GHz (22; 43 GHz) was used for all images and is shown in the lower left corner of each image sequence. Color images (on a linear scale shown at the top of each sequence) show the linearly polarized intensity. White bars (of unit length) indicate the electric vector position angle, {\it uncorrected} for Faraday rotation.}
\label{bg113}
\end{figure*}

  Observations during 2001 were made in dual polarization at the standard frequencies of 15, 22, and 43 GHz, covering a total of 12 epochs: 2000 December 30 (hereafter epoch A), 2001 February 1 (B), 2001 March 5 (C), 2001 April 1 (D), 2001 May 7 (E), 2001 June 8 (F), 2001 july 7 (G), 2001 August 9 (H), 2001 September 9 (I), 2001 October 11 (J), 2001 November 10 (K), and 2001 December 13 (L). The data were recorded in 2-bit sampling VLBA format with 32 MHz bandwidth per circular polarization at 256 Mbits/s, except for epochs B, C, D, F, and G, for which 1-bit sampling and 128 Mbits/s were used. One of the Very Large Array (VLA) antennas was added to the array at epochs A and E. Bad weather conditions and/or hardware problems, resulting in the loss of data, occurred at: HN during epochs A, C and D; OV for epoch H; and KP, MK and BR at epoch L.
  
  The 2007 observations were performed on November 3 (at 86 GHz), November 7 (43, 22, and 15 GHz), and November 30 (12, 8, 5, 2, and 1.7 GHz) with all available VLBA antennas except Saint Croix, which was down for maintenance. The data were recorded in 2-bit sampling VLBA format with 32 MHz bandwidth per circular polarization at a recording rate of 256 Mbits/s. Observations at 12 GHz were split into two 16 MHz bandwidths centered at 12.59 and 12.12 GHz to maximize possible detection of small Faraday rotation on relatively small angular scales.
  
  Reduction of the data was performed with the AIPS software in the usual manner \citep*[e.g.,][]{Leppanen:1995p1976}. Opacity corrections at 22, 43, and 86 GHz were introduced by solving for receiver temperature and zenith opacity at each antenna. The feed D-terms (instrumental polarization) were found to be very consistent over all observed sources and to remain stable across epochs during the 2001 observations.

\begin{figure*}
\epsscale{1.0}
\plotone{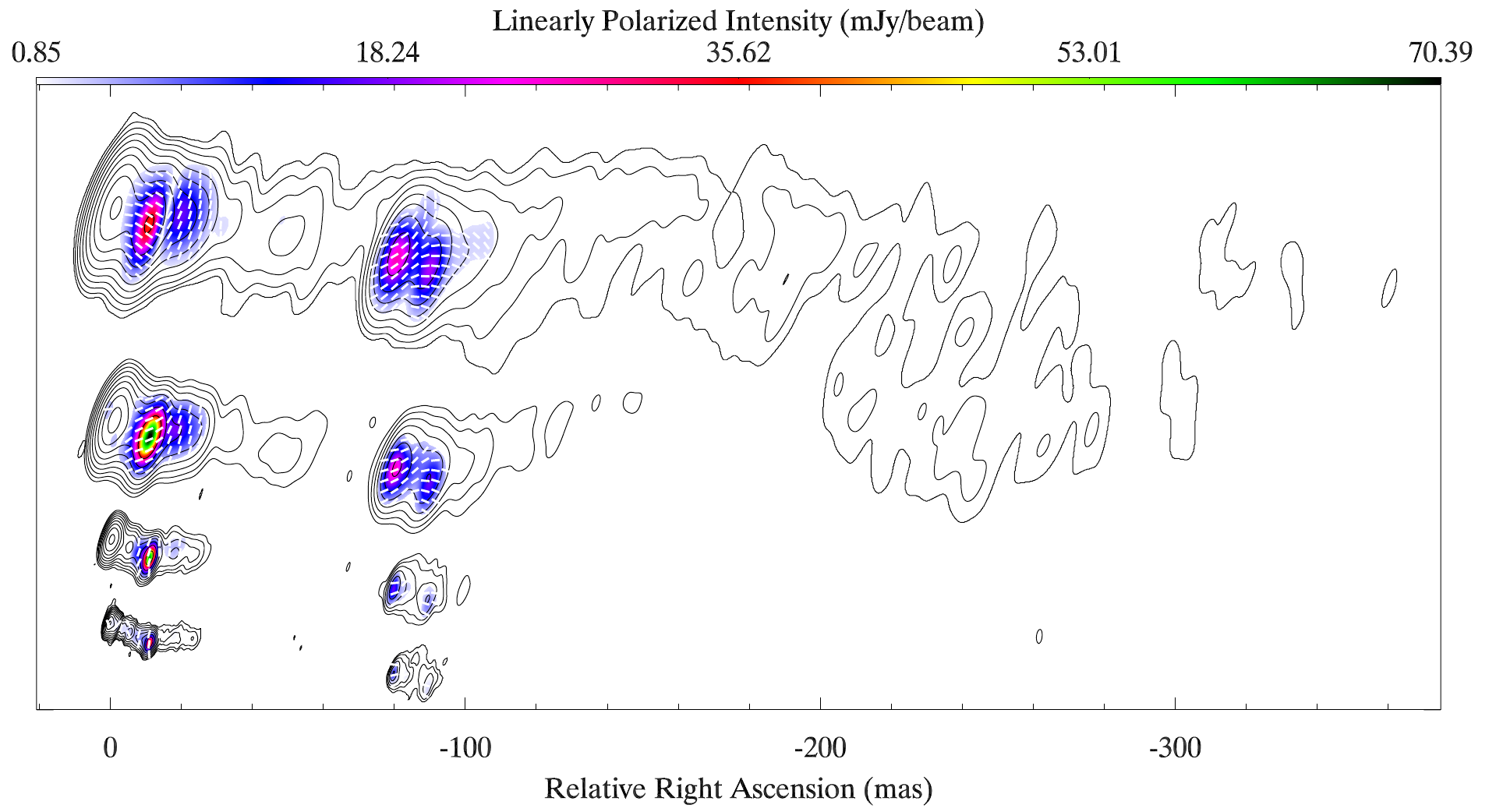}
\caption{VLBA images of 3C~120 in 2007 November 30 (2007.91) at 1.7, 2, 5 and 8 GHz, from top to bottom, respectively. Total intensity (naturally weighted) contours are overlaid at 0.9, 2.0, 4.2, 8.9, 19, 41, 86, 184, 392, and 834 mJy beam$^{-1}$, with convolving beams of FWHM 16.85$\times$5.55, 11.85$\times$4.11, 5.8$\times$1.9, and 3.36$\times$1.11 mas at -18$^{\circ}$ for frequencies 1.7, 2, 5, and 8 GHz, respectively. Colors show the linearly polarized intensity, and bars (of unit length) indicate the EVPAs, {\it uncorrected} for Faraday rotation.}
\label{1p7_8}
\end{figure*}

\begin{figure*}
\epsscale{1.0}
\plotone{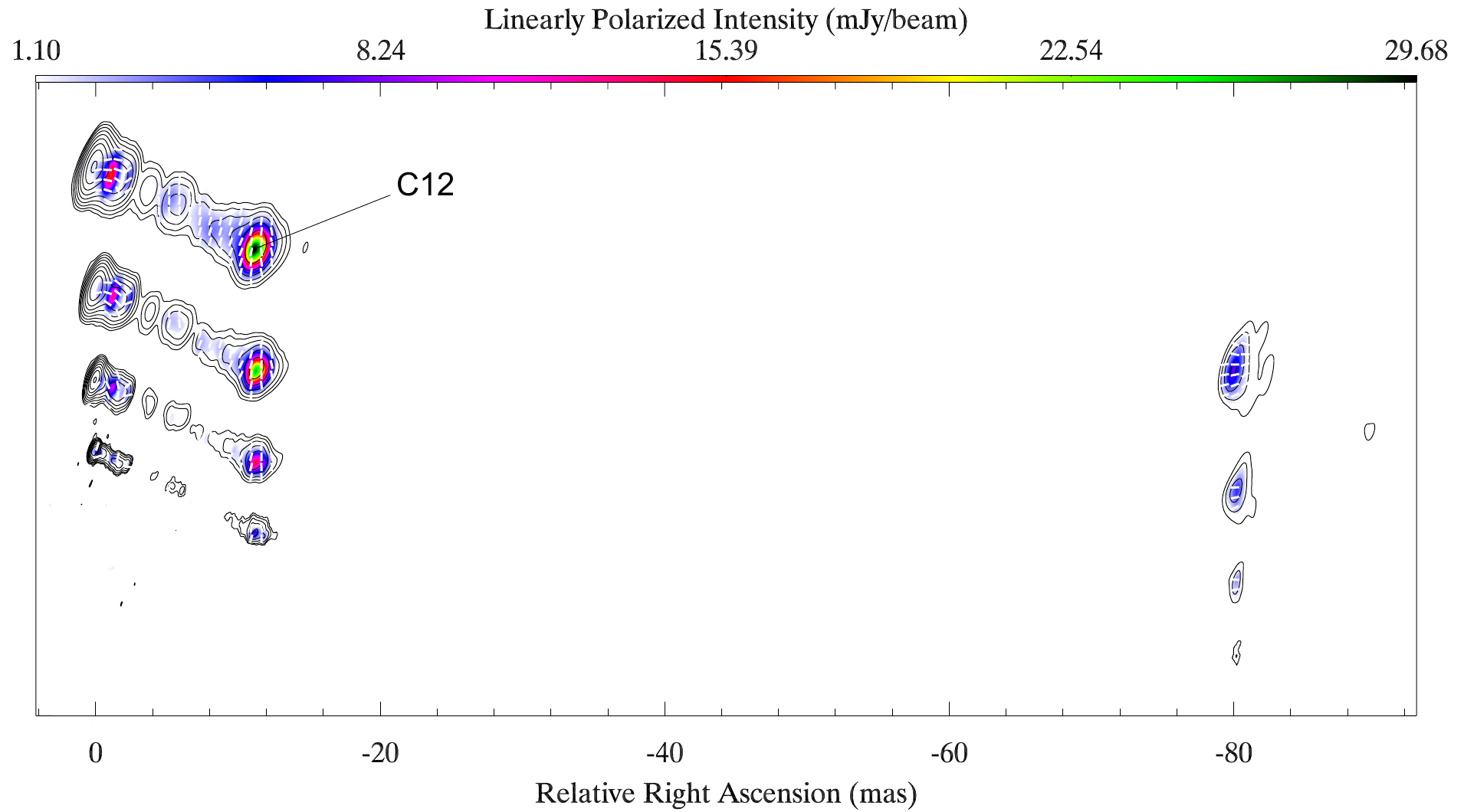}
\caption{Same as Fig.~\ref{1p7_8} for observations at 12 (2 IFs centered at 12.59 GHz), 15, 22, and 43 GHz, from top to bottom, respectively. Observations were taken in 2007 November 7 (2007.85), except at 12 GHz (epoch 2007.91). Contours are overlaid at 1.1, 2.4, 5.3, 11, 25, 53, 115, 249, 538, and 1164 mJy beam$^{-1}$, with convolving beams of FWHM 2.21$\times$0.71, 1.76$\times$0.61, 1.25$\times$0.4, and 0.63$\times$0.22 mas at -18$^{\circ}$ for frequencies 12, 15, 22, and 43 GHz, respectively.}
\label{12_43}
\end{figure*}

  The absolute phase offset between the right- and left-circularly polarized data, which determines the electric vector position angle (EVPA), was obtained by comparison of the integrated polarization of the VLBA images of several calibrators (0420$-$014, OJ~287, BL~Lac, 3C~454.3, DA193, 3C~279, and 4C~39.25) with VLA observations at epochs 2000 December 31, 2001 February 3, 2001 March 4, 2001 April 6, 2001 May 11, 2001 June 10, 2001 August 12, 2001 September 10, 2001 October 12, 2001 November 6, 2001 December 15, and 2007 November 19, as well as archival data from the UMRAO, MOJAVE, and NRAO long term monitoring programs. Estimated errors in the orientation of the EVPAs vary from epoch to epoch and with frequency, but usually lie in the range of 5$^{\circ}$-10$^{\circ}$. After the initial reduction, the data were edited, self-calibrated, and imaged both in total and polarized intensity with a combination of AIPS and DIFMAP \citep{Shepherd:1997p9326}. Comparison of the D-terms across epochs provides an alternative calibration of the EVPAs \citep{2002.VLBA.SM.30}, which was found to be consistent with that obtained by comparison with the VLA data.

\subsection{2001 images}

Figure \ref{bg113} shows the total and linearly polarized intensity images corresponding to the 2001 monitoring program. These were used to obtain the sequence and mean value of the rotation measure maps shown in Figs.~1 and 2 of \cite{Gomez:2008p1527}, respectively, which requires careful alignment of the images across epochs and frequencies. A detailed description of the procedure used and accuracy achieved is presented in Appendix \ref{img-alignment}.

\subsection{2007 images}

\begin{figure}
\epsscale{1.2}
\plotone{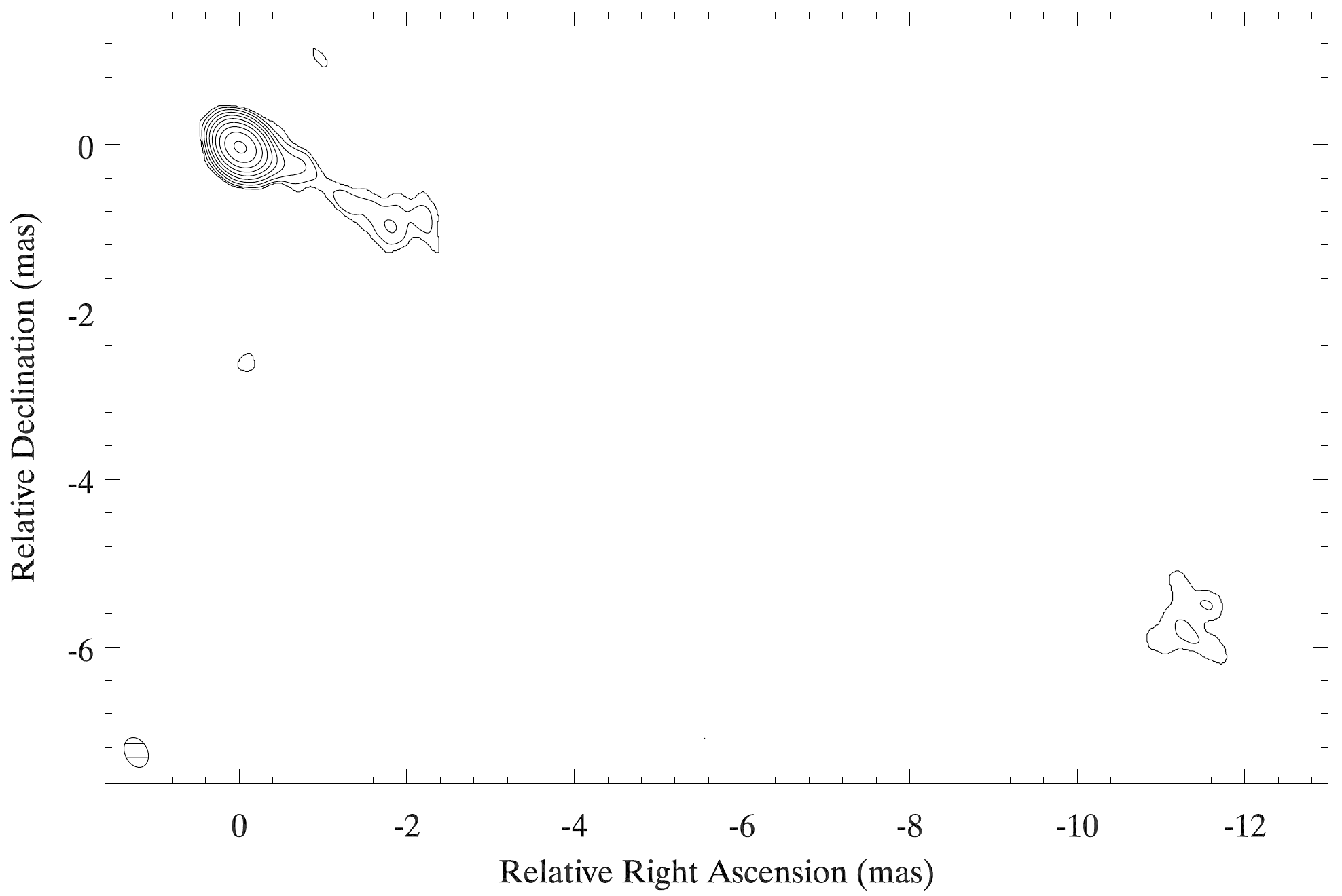}
\caption{Total intensity (naturally weighted) VLBA image of 3C~120 in 2007 November 3 (2007.84) at 86 GHz. Ten contours in equally spaced logarithmic intervals are plotted between 0.5 and 90\% of the peak brightness of 0.54 Jy beam$^{-1}$. The convolving beam of FWHM 0.37$\times$0.27 mas at 25$^{\circ}$ is shown in the lower left corner. No polarization was detected at this frequency.}
\label{86}
\end{figure}

  VLBA observations of 3C~120 taken during November 2007 are shown in Figs.~\ref{1p7_8}, \ref{12_43} and \ref{86}, which provide a 50-fold range in angular resolution. In \cite{RocaSogorb:2010p13195} we carried out a partial analysis of these data that revealed a brightness temperature for the component located 80 mas from the core about 600 times higher than expected at such distances. \cite{RocaSogorb:2010p13195} concluded that, although a helical shocked jet model may explain this unusually high brightness temperature, it appears that some other intrinsic process capable of producing a local burst in particle and/or magnetic energy may be needed to explain also the sudden appearance and apparent stationarity of this component.

  The 1.7 GHz total intensity image of Fig.~\ref{1p7_8} shows a continuous jet that extends approximately 300 mas, with a bulge of emission between 200 and 280 mas. Although the complex structure of the bulge prevents us from obtaining an accurate measure of its position, comparison by eye between our image and those by \cite{Walker:2001p2273} -- covering in total a time span of almost 30 years -- confirms the proper motion of 1.7$\pm$0.4 mas yr$^{-1}$ for the bulge reported by \cite{Walker:2001p2273}. This is similar to the proper motions measured at parsec scales \citep[e.g.,][]{Gomez:2001p1526}, and indicates that relativistic bulk velocities are sustained in the jet over at least hundreds of parsecs from the core.

  The progressively higher frequency images of Figs.~\ref{1p7_8} and \ref{12_43} reveal a rich structure in both total and linearly polarized intensity even at the shortest wavelengths, being the emission at 80 mas particularly remarkable, as reported by \cite{RocaSogorb:2010p13195}. The component 12 mas from the core (hereafter C12) is also especially bright. \cite{RocaSogorb:2010p13195} estimated a compression factor\footnote{Defined as the ratio of the upstream to downstream particle densities \citep{Hughes:1989p1865}.} $\eta\sim0.35$ assuming that the feature corresponds to a moving shock wave. This allows the detection of this component even at 86 GHz (Fig.~\ref{86}), at which some emission from the innermost 2 mas is also visible in the image. No polarization is detected at 86 GHz though. Comparison of the core flux density among frequencies yields a flat spectrum at 43 GHz and frequencies below, but optically thin at 86 GHz; we should note, however, that this may be affected by the relatively large uncertainties involved in the flux calibration at this frequency. Previous 86 GHz observations of 3C~120 established an upper limit to the overall core size of 54 $\mu$as \citep{Gomez:1999p1485}, followed by a measurement of the FWHM size of the core along the axial direction of $22\pm2$ $\mu$as, as well as an estimation of the number density of the combined electron-positron population in the core \citep{Marscher:2007p2015}.

\section {Faraday rotation analysis}

\subsection{Rotation measure images from 1999, 2001, and 2007}

  Th first indications of the presence of the localized Faraday rotation region between 3 and 4 mas from the core were obtained through monthly polarimetric VLBA observations at 22 and 43 GHz carried out between November 1997 and March 1999 \citep{Gomez:2000p1484,Gomez:2001p1526}. The EVPAs of several components were observed to rotate while they passed through this region, by an amount which followed a quadratic dependence with observing wavelength -- as expected for Faraday rotation. This is more clearly seen in the rotation measure map for epoch 10 January 1999 shown in Fig.~\ref{RMJan99}, obtained from observations at 22 and 43 GHz. To solve for the $\pm\pi$ ambiguities in the EVPAs when obtaining the RM image we have considered those values that result in the smallest RM values and yield the best agreement (both in RM and RM-corrected EVPAs) with the image obtained for observations in 2001 \citep[see Fig.~2 of][]{Gomez:2008p1527}, also plotted in Fig.~\ref{RMJan99} for comparison. The 2001 RM image reveals the presence of an RM screen in 3C~120 whose properties remain constant over one year, during which most of the jet shows rotation measure and RM-corrected EVPAs values with standard deviations smaller than 1000 rad m$^{-2}$ and 20$^{\circ}$, respectively. The Faraday screen displays a localized region of enhanced RM between approximately 3 and 4 mas from the core, with a peak of $\sim6000$ rad m$^{-2}$, as well as gradients across and along the jet.

  Rotation measure images for observations in 2007 combining the high, intermediate, and low frequency intervals are shown in Figs.~\ref{rm_15_43}, \ref{rm_5_12} and \ref{rm_1p7_8}, respectively. 

\subsection{Uncorrelated changes in jet field and Faraday screen 1999-2001}

\begin{figure}
\epsscale{1.2}
\plotone{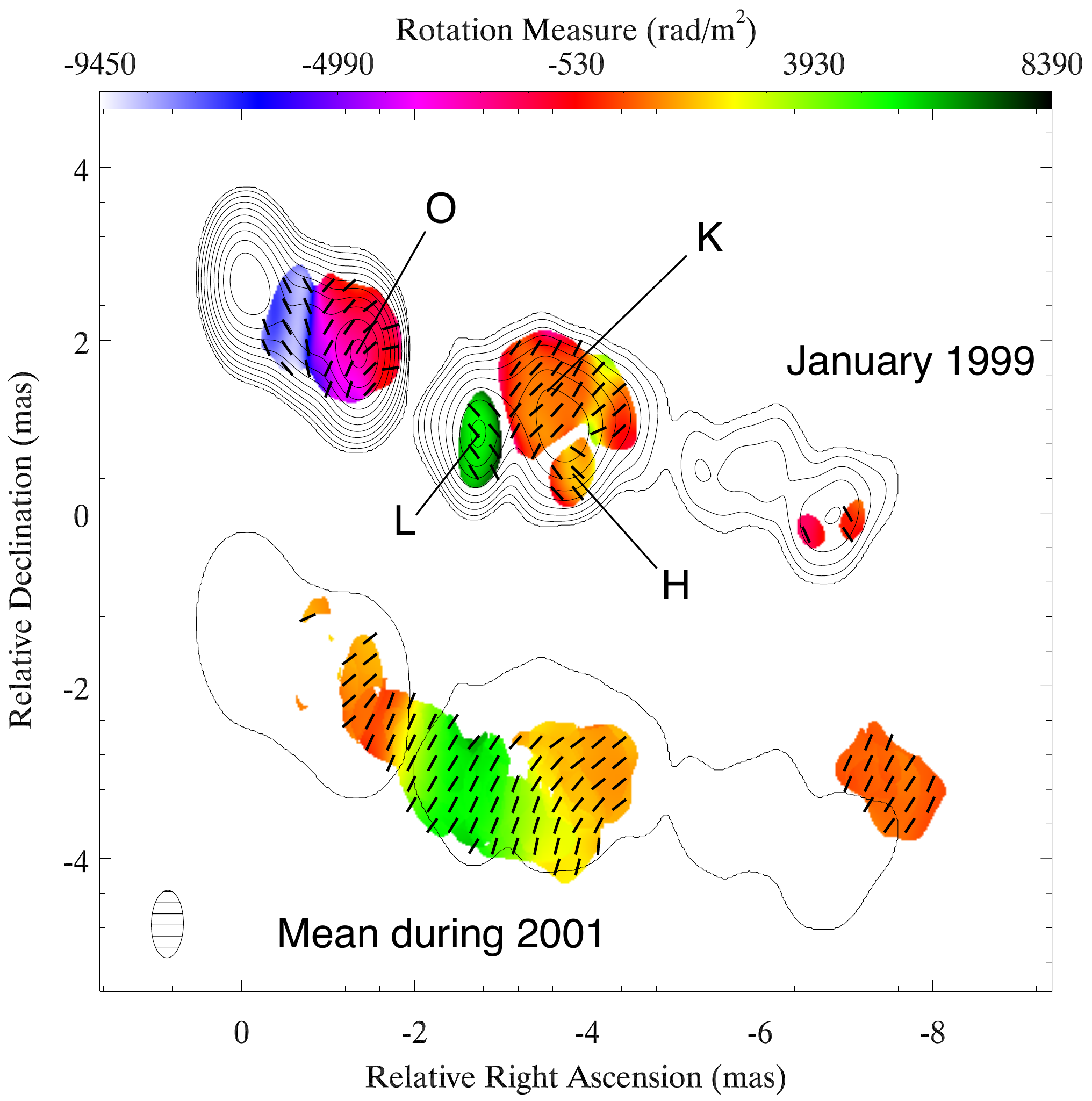}
\caption{({\it top}) Rotation measure map of 3C~120 in 10 January 1999 from VLBA observations at 22 and 43 GHz. Total intensity (naturally weighted) 22 GHz contours are overlaid at 3.6, 5.9, 9.8, 16, 27, 44, 73, 120, 198, 327 and 540 mJy beam$^{-1}$. {\it (bottom)} Map of the mean value of the rotation measure for observations during 2001 \citep[see also][]{Gomez:2008p1527}, shown for comparison. Bars (of unit length) indicate the RM-corrected electric vector position angle. The lowest contour of the January 1999 image is also shown to facilitate comparison.}
\label{RMJan99}
\end{figure}

\begin{figure}
\epsscale{1.2}
\plotone{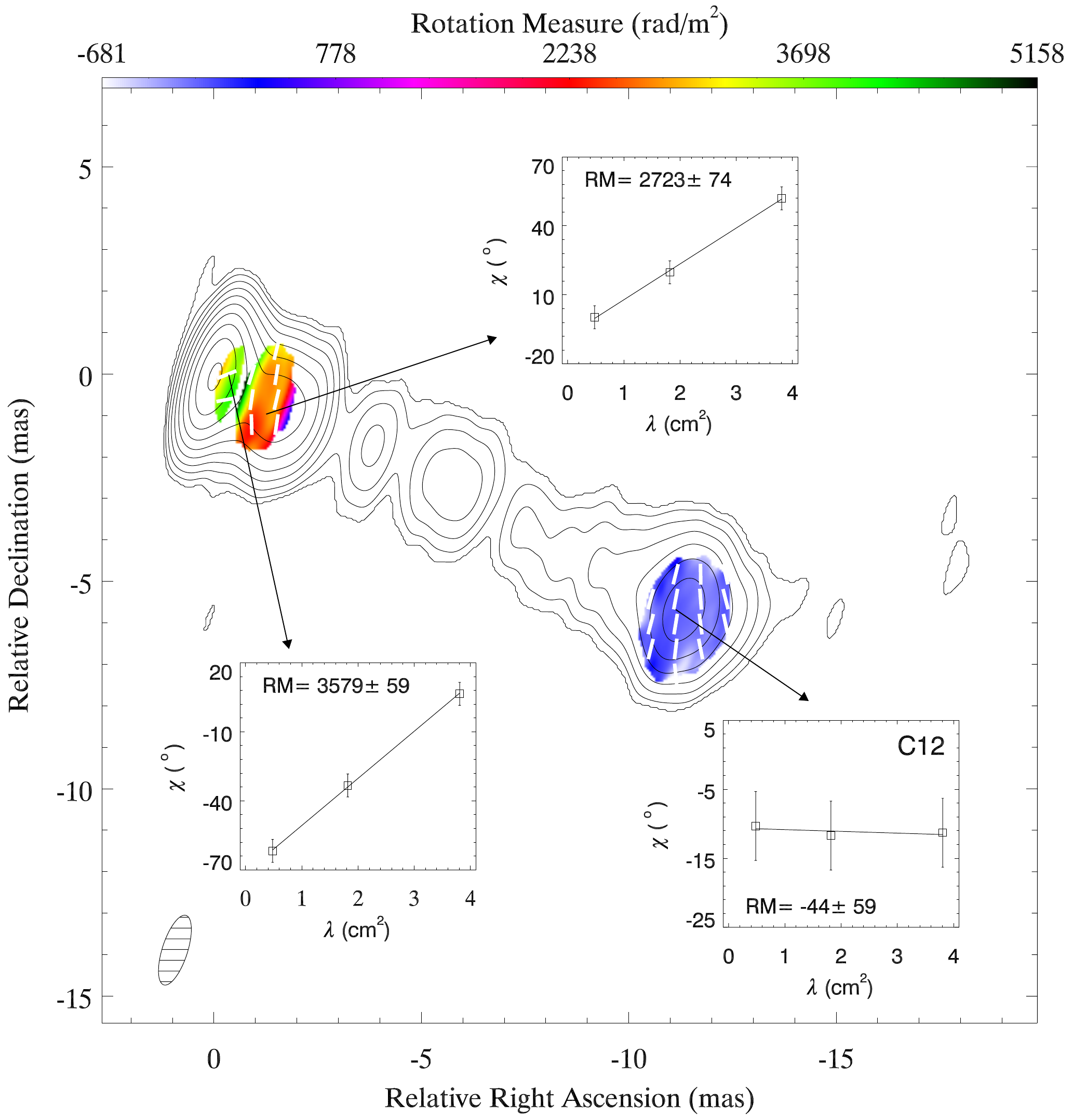}
\caption{Rotation measure image of 3C~120 between 15 and 43 GHz. Contours show the 15 GHz total intensity image. Bars indicate the RM-corrected EVPAs. Inset panels show sample fits to a $\lambda^2$ law of the EVPAs at some particular locations. The convolving beam is shown in the lower left corner.}
\label{rm_15_43}
\end{figure}

  To quantify any possible variation in the Faraday screen, we have subtracted the RM values of January 1999 from those of the mean RM map for 2001, and computed the mean value of the difference in the components (jet areas in the mean 2001 map) labeled in Fig.~\ref{RMJan99} \citep[following the nomenclature of][]{Gomez:2000p1484,Gomez:2001p1526}. Component L is located at the localized region of high Faraday rotation, which is found to remain in the same jet area as in 2001, with very similar values of the rotation measure (mean difference of $-660\pm770$ rad m$^{-2}$). However, the RM-corrected EVPAs show significant rotation, with a mean difference between both epochs of $-64\pm9^{\circ}$. A similar situation is found for component H, with a mean variation in the RM of $1030\pm1010$ rad m$^{-2}$ and a rotation of the RM-corrected EVPAs of $-72\pm19^{\circ}$. On the other hand, component K shows no significant changes in either the RM-corrected EVPAs or RM, with variations of $-3\pm19^{\circ}$ and $810\pm930$ rad m$^{-2}$, respectively. For component O we also find very similar RM-corrected EVPAs, with a variation of $8\pm28^{\circ}$. However, the rotation measure in O is observed to have changed significantly, with a difference between both epochs of $2730\pm910$ rad m$^{-2}$.
  
  Therefore, we find uncorrelated changes in the linear polarization of the underlying jet emission and the Faraday rotation screen: while the RM remains constant in the outer jet -- including the localized region of high RM -- the RM-corrected EVPAs of two particular components (L and H) rotate by almost $90^{\circ}$; on the other hand, in the innermost 2 mas the RM changes significantly but without variations in the RM-corrected EVPAs. These uncorrelated changes suggest that the emitting jet and source of RM are not closely connected physically.

  Components L and H have RM-corrected EVPAs closely aligned with the jet, in agreement with observations by \cite{Jorstad:2005p1946}. This is contrast to what we find during 2001, where the RM-corrected EVPAs are observed to be approximately perpendicular to the jet direction. This suggests an aligned magnetic field, in agreement with that observed on larger scales \citep{Walker:1987p2272}. If these components correspond to strong plane-perpendicular shock waves, compression of the magnetic field parallel to the shock front can result in a dominant magnetic field perpendicular to the components' motion \citep[e.g.,][]{Marscher:1985p2022,Hughes:1989p1865,Gomez:1994p1483}, as observed in knots L and H. However, no similar orientation of the EVPAs is found for component O -- which is probably also associated with an even stronger shock -- or in any of the other components that have traveled along the same jet regions as that of L and H during the 2001 observations \citep[see also][]{Jorstad:2005p1946}. On the other hand, if components L and H are not particularly strong shocks -- which finds support by the fact that their degrees of polarization are similar to those observed for other components at similar jet locations -- we can consider the possibility that these components either have an intrinsically different magnetic field (mainly perpendicular to the jet axis), or that they have a different velocity and/or orientation with respect to the observer. In this latter case, the Lorentz transformation of the magnetic field -- which amplifies the components transverse to the motion -- would generally result in an \textit{observed} dominant toroidal field for an assumed helical magnetic field \citep[e.g.,][]{Blandford:1979p1412,Lyutikov:2005p2007}. The clear stratification in polarization observed between the northern and southern jet regions \citep[see Fig.~3 of][]{Gomez:2008p1527} and the fact that both components L and H move along the southern side may support this idea of stratification across the jet of the magnetic field and/or kinematics.

\subsection{No detection of the localized high RM region 2007: an effect of sampling}

\begin{figure*}
\epsscale{1}
\plotone{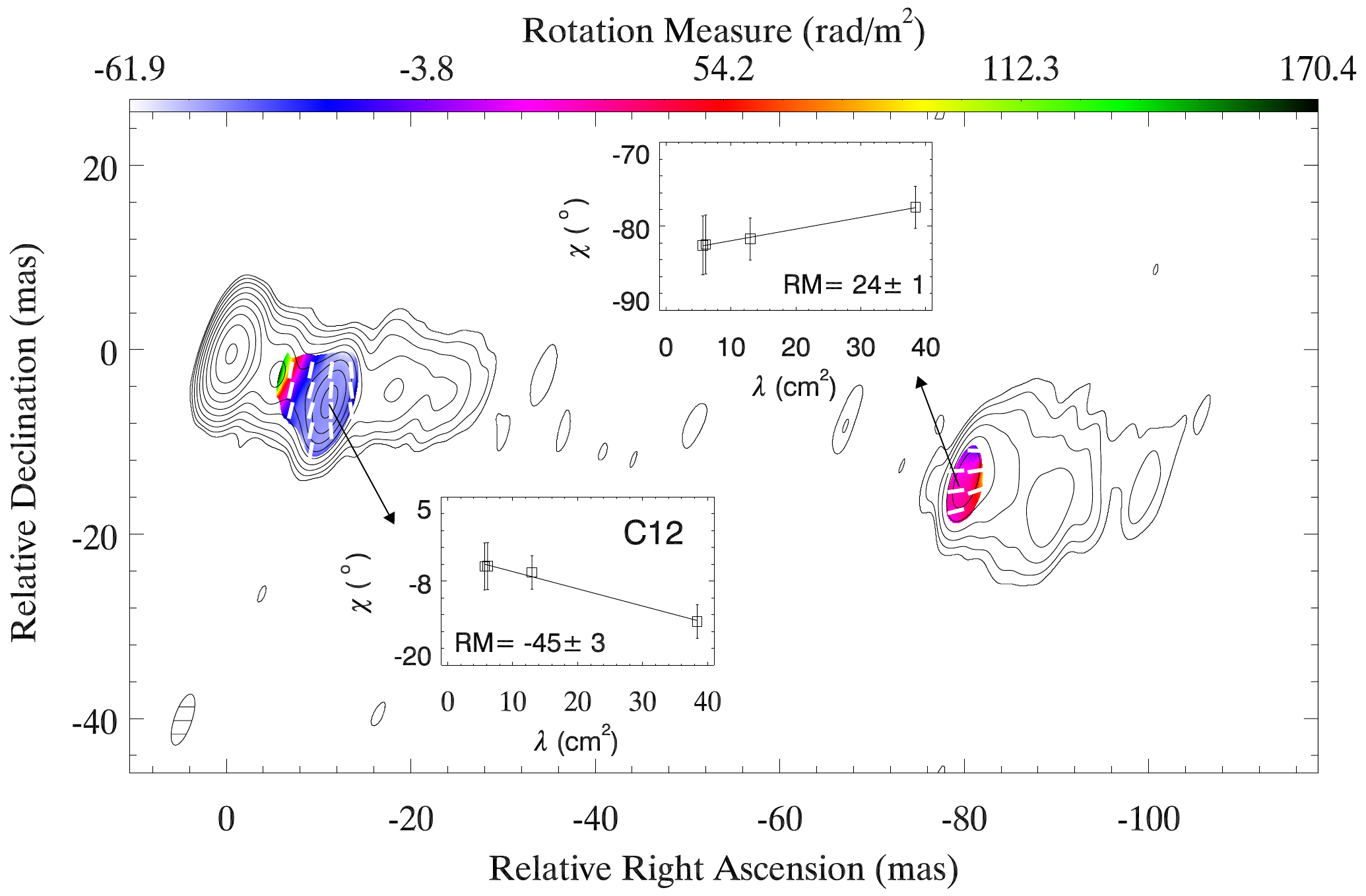}
\caption{Same as Fig.~\ref{rm_15_43} for frequencies between 5 and 12 GHz. Contours show the 5 GHz total intensity image.}
\label{rm_5_12}
\end{figure*}

\begin{figure*}
\epsscale{1}
\plotone{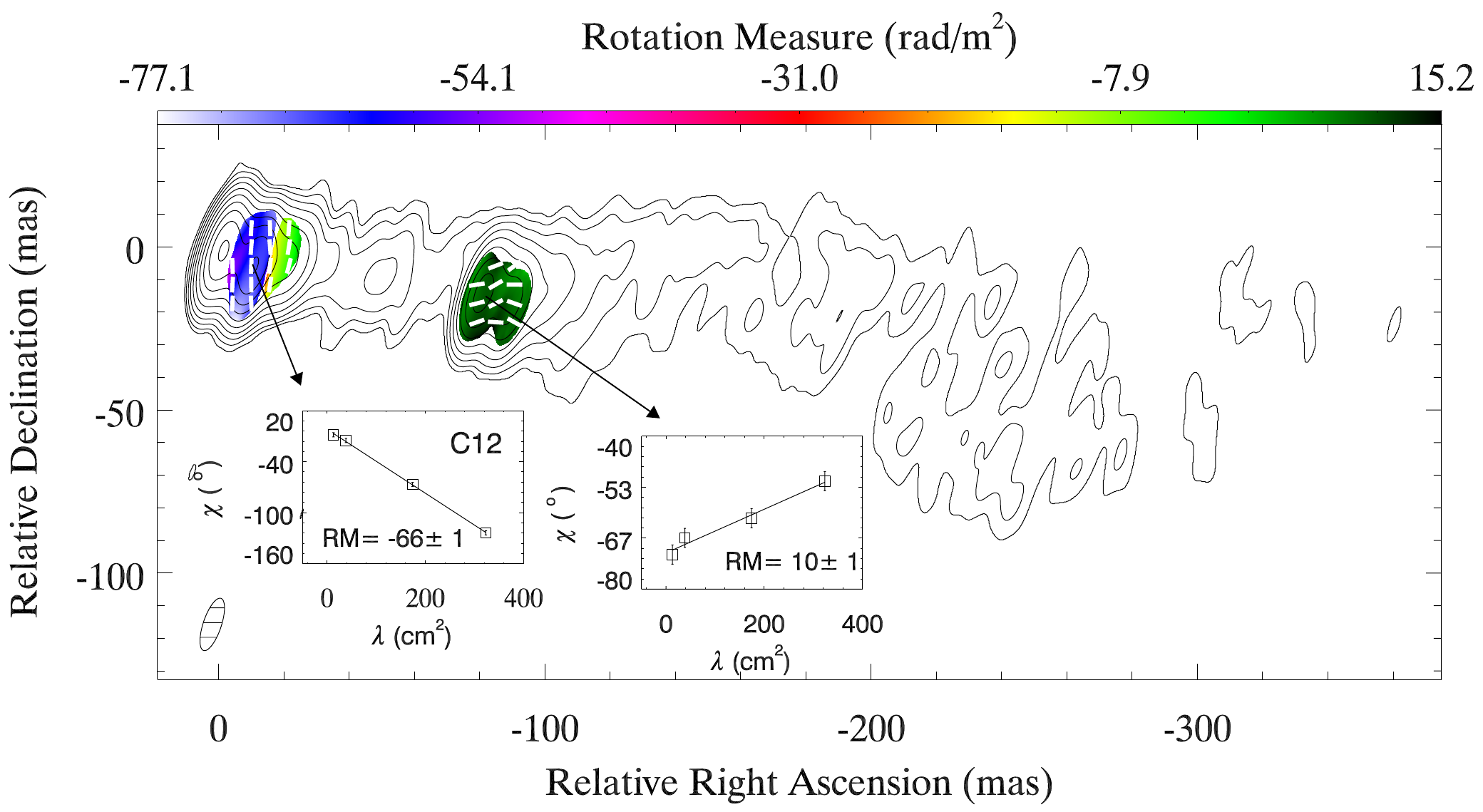}
\caption{Same as Fig.~\ref{rm_15_43} for frequencies between 1.7 and 8 GHz. Rotation measure image of 3C~120 between 1.7 and 8 GHz. Contours show the 1.7 GHz total intensity image.}
\label{rm_1p7_8}
\end{figure*}

\begin{figure}
\epsscale{1.15}
\plotone{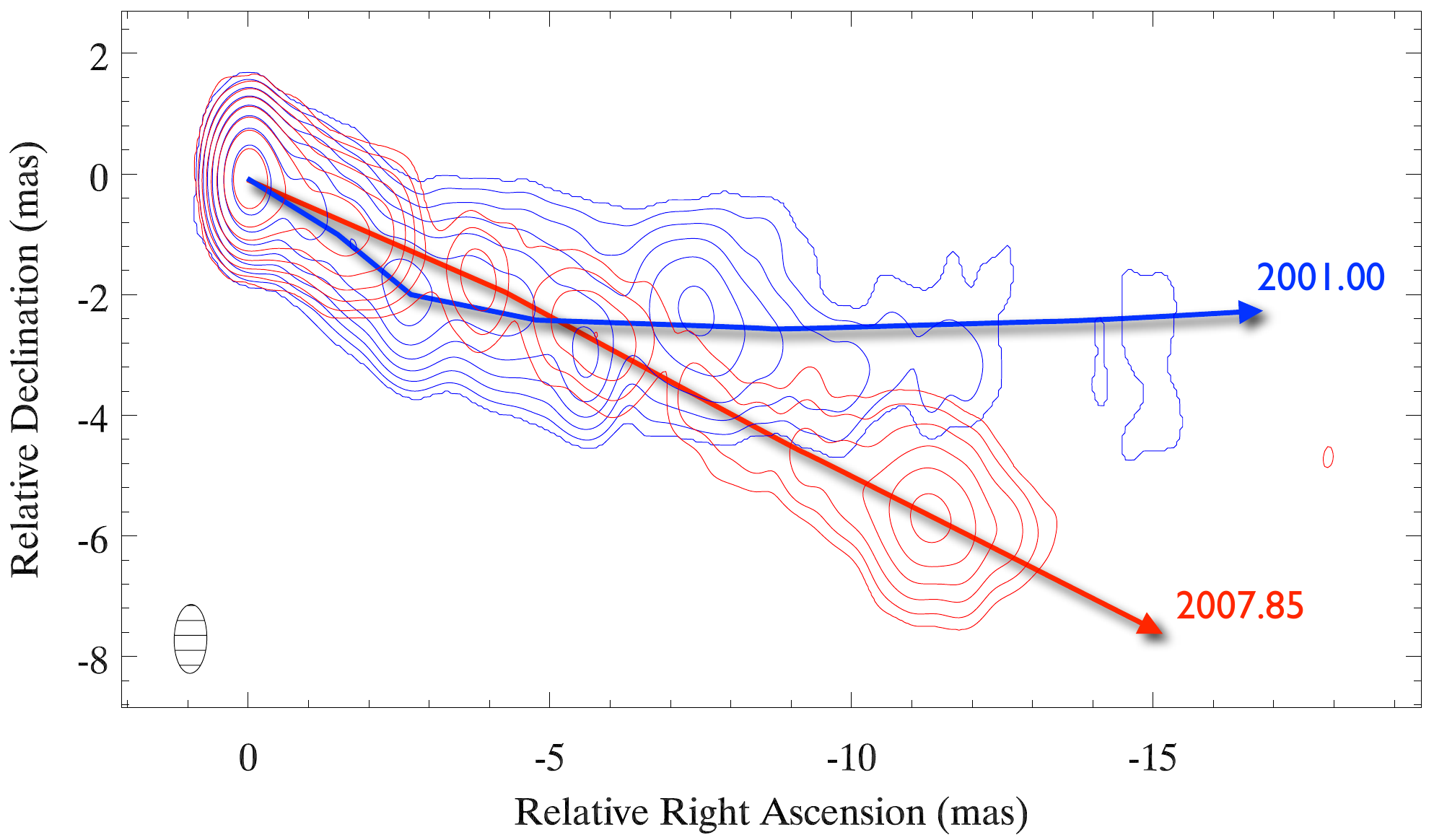}
\caption{Total intensity VLBA images of 3C~120 for epochs 2001.00 (blue) and 2007.85 (red) at 15 GHz. Both images have been convolved with the same beam of 1.14$\times$0.54 mas at -1$^{\circ}$ (shown in the lower left corner) for easier comparison.}
\label{tf01_07}
\end{figure}

  The middle and low frequency range RM maps (Figs.~\ref{rm_5_12} and \ref{rm_1p7_8}) reveal no indication of localized regions of high rotation measure at these scales similar to that found at parsec scales in previous observations \citep[][see also Fig.~\ref{RMJan99}]{Gomez:2008p1527}, as may have been expected given the wealth of evidence for interaction between the jet and ambient medium \citep[e.g.,][]{Axon:1989p7975,Sanchez:2004p2244,Gomez:2000p1484,Gomez:2008p1527}. The high frequency range RM map of Fig.~\ref{rm_15_43} does \textit{not} show the localized region of high RM that was found in previous observations. This can be understood by examining Fig.~\ref{tf01_07}, which shows the superposition of the 15 GHz total intensity images at the 2001.00 and 2007.85 epochs, revealing a significant change in the jet geometry with time. The position angle of the innermost 3 mas structure has changed from $-126^{\circ}\pm2$ in 2001.00 to $-114^{\circ}\pm2$ in 2007.85; that is, the jet direction of ejection rotated to the north by $12^{\circ}$ between these two epochs. As a consequence of this rotation the components in the 2007 jet do not travel across the localized region of enhanced RM. They thus fail to reveal it, since it is only through the motion of superluminal components that we are able to map the jet polarization owing to the increase in energy density and magnetic field ordering that they produce.
  
  The rotation of the direction of ejection is in agreement with the helical structure suggested by \cite{Gomez:2001p1526}, and the proposed precessing/helical models of \cite{Caproni:2004p1426} and \cite{Hardee:2005p1755}. In particular, \cite{Caproni:2004p1426} estimated a precession period of 12.3 yr, predicting a swing to the north of the direction of ejection in the jet between our 2001 and 2007 epochs, as is indeed observed.

  There is evidence for non-ballistic trajectories of the superluminal components for observations between November 1997 and March 1999 \citep{Gomez:2000p1484,Gomez:2001p1526}, resulting in the gentle bend toward the north observed in the 2001 epoch of Fig.~\ref{tf01_07}. This contrasts with the rectilinear structure observed at the 2007.85 epoch. Component C12 is one of the brightest ever observed at parsec scales in 3C~120, so it is possible that it contains an unusually large energy density, with enough momentum to move ballistically and "break through" the bent jet funnel inferred in the 2001 epoch. This may have left behind a funnel through which upstream components move ballistically, resulting in the almost rectilinear jet observed in 2007.85.

  Similar changes in the geometry of the innermost structure of the jet have also been found in the BL Lac object PKS~0735+178. In this case, they are related to changes in the overall activity of the source: quasi stationary or subluminal components moving through a twisted structure are observed during periods of low activity, while superluminal components moving along a straight jet are found during periods of high activity \citep{Gomez:1999p1488,Gomez:2001p1528,Gabuzda:2001p1461,Agudo:2006p1392,Britzen:2010p13518}.
  
  If component C12 is interacting with the external medium, it does not result in enhanced rotation measure. Figures \ref{rm_15_43} and \ref{rm_5_12} show that the rotation measure of C12 is approximately $-45$ rad m$^{-2}$ ($-66\pm1$ rad m$^{-2}$ at the lower frequency interval of Fig.~\ref{rm_1p7_8}), significantly smaller than the $\sim6000$ rad m$^{-2}$ found in the region between 3 and 4 mas from the core \citep[][see also Fig~\ref{RMJan99}]{Gomez:2008p1527}, presumably produced by the interaction of the jet with the external medium. However, there is a sign reversal in the rotation measure of C12 when compared with the remainder of the jet.

\subsection{RM on larger scales from 2007 and evidence for reversals}
  
  A progressive decrease in the RM along the jet in 3C~120 is visible in the 2007 observations (Figs.~\ref{rm_15_43}, \ref{rm_5_12}, and \ref{rm_1p7_8}), as well as in the January 1999 image of Fig.~\ref{RMJan99}. The relatively high RM close to the core may be responsible for the dramatic decrease in the degree of polarization toward the core, as previously suggested by \cite{Gomez:2008p1527}. Similar decreases in the RM along the jet have been previously observed in other sources \citep[e.g.,][]{Taylor:1998p14529,Gabuzda:2003p1456,Jorstad:2007p1952,OSullivan:2009p8966}. This can be explained as the result of the decrease in the density of the external ionized gas and magnetic field strength with distance from the core \citep[e.g.,][]{Guidetti:2010p13538}.
  
  Component C12 departs from this monotonic decrease in RM with core distance, undergoing a change in the sign of the RM, which implies a double sign reversal when analyzing the RM along the remainder of the jet. We note, however, that our RM images -- and therefore the existence of sign reversals -- may be affected by a source of RM in the direction of 3C~120 that is unrelated to the jet, which recent observations estimate to be $\sim$23 rad m$^{-2}$, with a 1-$\sigma$ error of $\sim$10 rad m$^{-2}$ \citep{Taylor:2009p14909}.

  RM sign reversals have been previously reported along \citep{Zavala:2001p2284,OSullivan:2009p8966} and across \citep{Gabuzda:2004p1462,Asada:2008p6013,Mahmud:2009p11368} jets, but to our knowledge, no similar double sign reversal at parsec scales as reported here has been found previously \citep[see also][]{Guidetti:2010p13538}. Sign reversals in the RM across a jet can be readily explained by a sheath threaded by a helical magnetic field. To account for the reversals along a jet, different alternative explanations have been proposed. One is based on ``magnetic tower" models, in which the magnetic field lines are wound up in two nested helices with opposite poloidal polarities, with the inner helix less tightly wound than the outer helix. Such magnetic towers are expected to arise by the Pointing-Robertson effect \citep{Contopoulos:1998p14824}, caused by the differential rotation of the accretion disc. A change in the relative contributions to the observed RM of the inner and outer helical fields has been considered by \cite{Mahmud:2009p11368} to explain changes in the direction of transverse RM gradients, which can eventually also lead to a sign reversal. Similar magnetic tower models have also been considered to explain the different helical pitch angles derived for the emitting jet and the sheath in NRAO 140 \citep{Asada:2008p6013} and S5 0836+710 \citep{Asada:2010p13535}. It is also possible to explain a sign reversal in the RM along the jet by assuming that the sheath responsible for the Faraday rotation is at least moderately relativistic, so that light aberration can lead to a change in the line-of-sight dominant component of the magnetic field through a small change in the direction or velocity of the flow, as considered previously by \cite{OSullivan:2009p8966}.

  It would be difficult to explain the double RM sign reversal along the jet of 3C~120 trough magnetic tower models, since this would consequently require an unlikely double change in the relative contributions to the RM of the inner and outer helical fields. However, the change in the RM sign in C12, when compared with the RM of the remaining jet, can be explained within the moderately relativistic sheath model as produced by a local change in the velocity or orientation, as would be expected in the case of precession in 3C~120. Note that this could also explain its unusually bright emission through differential Doppler boosting.
  
  It is interesting to note that, although the RM of C12 undergoes a sign reversal, the RM-corrected EVPAs have the same orientation as for the remainder of the optically thin jet, i.e., perpendicular to the jet. \citep[Note that component C80 contains RM-corrected EVPAs aligned with the jet, but as reported by][this is probably due to its peculiar nature.]{RocaSogorb:2010p13195} Therefore, in order to explain the RM sign reversal in C12 through any of the different proposed models without producing a change in the RM-corrected EVPAs, we need to assume that the emitting jet and the sheath responsible for the Faraday screen have different kinematical properties and/or magnetic field configurations. Alternatively, a much simpler interpretation for the different RM sign in C12 can be obtained by considering that the Faraday rotation is produced in a foreground cloud with a different magnetic field than that of the remaining jet.
  
  A sign reversal in the rotation measure along the jet is also observed in the January 1999 image (Fig.~\ref{RMJan99}), as well as between the innermost 2 mas of the jet in this image and those obtained in 2001 \citep[Fig.~1 in][]{Gomez:2008p1527} and 2007.85 (Fig.~\ref{rm_15_43}). If we assume that the sheath is at least moderately relativistic, a change in the direction of ejection in the jet -- as expected for precession -- could explain the differences in the RM between 1999 and 2007 through the different projection of the field along the line of sight. To explain the RM sign reversal between 1999 and 2001, an unlikely large change in the bulk velocity of the sheath would be needed (given the similar directions of ejection for the jet at both epochs), so that a change in the line-of-sight component of the field could be obtained through light aberration. Note however that, as mentioned for C12, different kinematical properties and/or magnetic field configurations would be needed for the sheath and emitting jet, so that the RM can change without producing a rotation of the RM-corrected EVPAs.
    
  \cite{Broderick:2009p11078} have considered the possibility of having relativistic \textit{helical} motions in the Faraday rotating sheath, showing that sign reversal in the rotation measure can naturally arise in this case, especially for observations at high frequencies that probe the innermost regions where there is a transition between ultra-relativistic and moderately relativistic helical motion. This could explain the RM sign reversals along the jet in the January 1999 image, but it would require a -- a priori unlikely -- considerable change in the kinematical properties of the sheath to account for the RM time variability observed between January 1999 and later epochs.
  
 The relativistic helical sheath model of \cite{Broderick:2009p11078} might also account for the sign reversal of component C12, which could also explain the variation in the RM with changing frequency intervals.

\section{Summary and conclusions}

  We have studied the source of the Faraday rotation screen in the radio galaxy 3C~120 through VLBA observations carried out between January 1999 and November 2007. Sampling of the rotation measure from the innermost regions up to hundreds of mas from the core is obtained in November 2007 through observations at 86, 43, 22, 15, 12, 8, 5, 2, and 1.7 GHz, providing a 50-fold range in angular resolution.

  By analyzing correlated motions of some particular superluminal components when performing the registration of images across epochs, we show that it is possible to achieve an accuracy between 12 and 30 $\mu$as, similar to what can be obtained by phase-reference VLBI astrometry.

  Significant changes are observed in the RM when comparing observations taken in 1999 and 2001 with those in 2007. In particular, one of the main features observed during the 1999 to 2001 observations, the existence of a localized, very stable high RM region between 2 and 3 mas from the core, is not observed. This is explained not as a result of an actual change in the Faraday screen, but by a transverse shift in the positions of bright features in the jet, such that the screen no longer covers superluminal knots 2-3 mas from the core. A similar explanation has been offered to explain the RM variability in 3C~279 and 3C~273 \citep{Taylor:2000p14342,Zavala:2001p2284}. The different trajectories of the superluminal components (i.e., different jet geometry) are in agreement with proposed precession models for 3C~120 \citep{Caproni:2004p1426,Hardee:2005p1755}.
  
  Comparison of observations taken between January 1999 and December 2001 reveals uncorrelated changes in the linear polarization of the underlying jet emission and the Faraday rotation screen: while the rotation measure remains constant between approximately 2 and 5 mas from the core during these three years of observations, the RM-corrected EVPAs of two particular superluminal components (moving along the southern side of the jet) are found to be rotated by almost $90^{\circ}$ when compared to other components moving through similar locations in the jet. On the other hand, in the innermost 2 mas the RM changes significantly -- including a sign reversal -- but shows no variations in the RM-corrected EVPAs. These uncorrelated changes suggest that the emitting jet and the source of the RM are not closely connected physically.
  
  Similarly, observations in 2007 reveal a double sign reversal in the RM along the jet while the RM-corrected EVPAs remain perpendicular to the jet axis. The variation in the RM can be accounted for by considering that the sheath responsible for the Faraday rotation is at least moderately relativistic, so that light aberration can lead to a reversal in the polarity of the line-of-sight magnetic field through a small change in the direction and/or velocity in the flow of the sheath. Note, however, that different kinematical properties and/or magnetic field configurations would be needed for the sheath and emitting jet, so that the RM can change without producing a rotation of the RM-corrected EVPAs. Such differences in magnetic field structure can be explained by a magnetic tower model, in which the inner helix would thread the emitting jet and the outer helix the RM sheath, but an unlikely double change in the ratio of the RM of the inner to the outer helix would be required to explain the double sign reversal in RM along the jet.
  
  The observed coherent structure and gradient of the RM along the jet supports the idea that the Faraday rotation is produced by a sheath of thermal electrons that surrounds the emitting jet. However, the uncorrelated changes in the RM screen and RM-corrected EVPAs suggest that the emitting jet and the source of Faraday rotation are not closely connected physically and have different configurations for the magnetic field and/or kinematical properties. Furthermore, the existence of a three-year-long stationary region of enhanced RM requires a localized source of Faraday rotation, which favors a model in which a significant fraction of the Faraday rotation measure found in 3C~120 originates in foreground clouds, rather than in a sheath intimately associated with the emitting jet. In this case, Faraday rotation studies will provide valuable information about the ambient medium through which jets propagate, but will not be able to reveal further details about the emitting jet, such as the line-of-sight magnetic field, and hence to test whether they are threaded by helical magnetic fields.

\acknowledgements This research has been supported by the Spanish Ministry of Science and Innovation grants AYA2007-67627-C03-03 and AYA2010-14844, by the Regional Government of Andaluc\'{\i}a (Spain) grant P09-FQM-4784, and by National Science Foundation grant AST-0907893. We thank the anonymous referee for helpful comments that improved our manuscript. The VLBA is an instrument of the National Radio Astronomy Observatory, a facility of the National Science Foundation operated under cooperative agreement by Associated Universities, Inc. This research has made use of data from the University of Michigan Radio Astronomy Observatory which has been supported by the University of Michigan and by a series of grants from the National Science Foundation, most recently AST-0607523.

{\it Facilities:} \facility{VLBA ()},\facility{VLA ()}

\appendix
\section{Registering of the 2001 images}\label{img-alignment}

\begin{figure*}
\epsscale{1.15}
\plotone{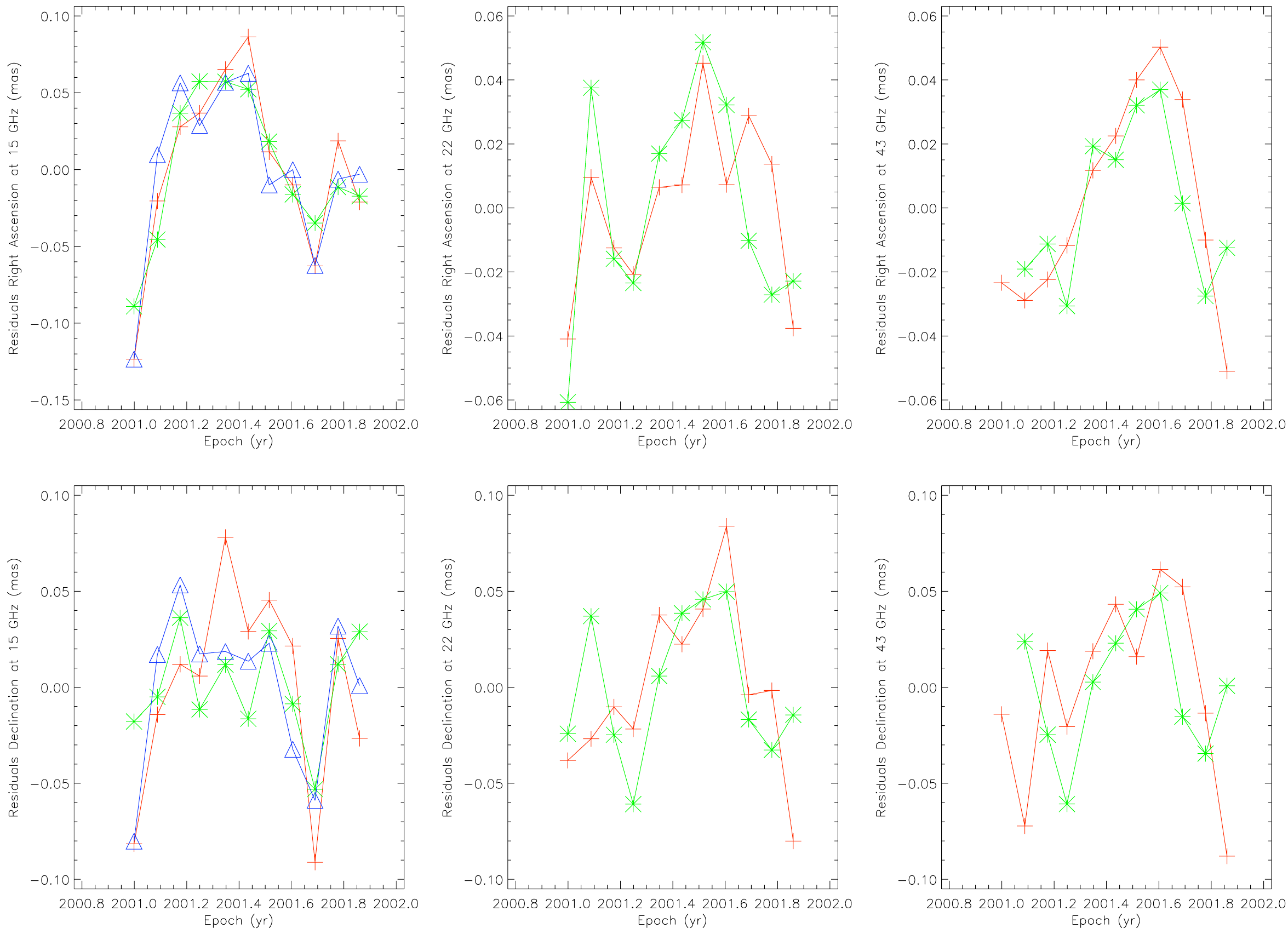}
\caption{Residuals of the linear fits in right ascension and declination with time for the registering components at 15 GHz (\emph{left}), 22 (\emph{middle}) and 43 GHz (\emph{right}). Blue triangles correspond to component \emph{b}, green asterisks to \emph{c} and red crosses to \emph{f}. Component \emph{c} appears blended with other jet features at 43 GHz in epoch A, and therefore has not been used in the analysis.}
\label{res}
\end{figure*}

  Analysis of the images corresponding to the 2001 monitoring (Fig.~\ref{bg113}) across epochs and frequencies requires first a proper alignment (or registration) of the maps, given that self-calibration of the visibility data during the imaging process results in the loss of absolute positioning of the source. To avoid this it is possible to perform phase-referenced observations, in which the position of the target source can be determined relative to a reference source. However, this requires the reference source to be strong, close to the target source, and ideally point-like. These conditions are rarely met, and therefore the alignment of VLBI images is commonly performed by looking for particular jet feature(s) that can be unambiguously identified in each of the images to be compared. The core (defined here as the position where the jet becomes optically thin) seems \emph{a priori} a good choice, since it is easy to identify. However, opacity effects can move its position downstream as the frequency of observation decreases \cite*[e.g.,][]{Blandford:1979p1412,Gomez:1993p1478,Lobanov:1998p9900,Caproni:2004p1426,Hirotani:2005p13720,Kovalev:2008p5779}, rendering the alignment of images at multiple frequencies a difficult task. Furthermore, the ejection of a new component produces a temporary dragging of the apparent core position downstream. This is due to the increased opacity and the motion of the peak brightness resulting from the convolution of the actual core and the new moving component. Variations in the jet flow velocity or viewing angle, as expected for precessing jets, should also lead to a change in the core position \citep[e.g.,][]{Caproni:2004p1426}. This jitter of the core position complicates the alignment of images, especially in active sources like 3C~120. It is therefore more convenient to choose components that are known to be optically thin, so that opacity effects do not affect the position when images are compared at different frequencies. To minimize the errors in the determination of the position such components are ideally strong and compact, that is, have high brightness temperatures. Finding such components is not always easy, since jet features can be observed to split or merge, or to be difficult to disentangle from other neighboring components. In an effort to overcome such difficulties, \cite{Croke:2008p1437} describe a new method based on image cross-correlation techniques, and show that it can be used to align maps at different frequencies. More recently, \cite{Dodson:2009p14083} have reported a novel phase-reference procedure, dubbed \textit{source/frequency phase referencing} (SFPR), which combines the source switching of conventional phase-referencing techniques with fast frequency switching, allowing astrometric measurements of frequency-dependent core positions up to 86 GHz.

  The characterization of the images of Fig.~\ref{bg113} in terms of discrete jet components has been carried out using \textsf{DIFMAP}, which allows one to fit the visibility data with Gaussian brightness distributions. Multiple jet components can be identified in the images of Fig.~\ref{bg113}. Most of these display complex evolution \cite*[see also, e.g.,][]{Gomez:2000p1484,Gomez:2001p1526,Homan:2001p1758,Jorstad:2005p1946,Agudo:2007p2404,Agudo:2010p14828}, or are difficult to match at the three observing frequencies simultaneously, and therefore cannot be used to register the images. However, components \emph{b}, \emph{c}, and \emph{f} at 15 GHz, and \emph{b} and \emph{f} at 22 and 43 GHz (see Fig.~\ref{bg113}) can be very reliably identified across epochs and frequencies, and therefore can be used to align the maps. We refer hereafter to these as the ``registering components".
  
  \textsf{DIFMAP} sets the coordinate origin at the phase center of the visibility data, that is, the peak in the initial map. This is always close to, but not necessarily coincident with, the peak brightness of the final clean image, which in turn usually coincides with the core as long as the jet is core dominated. That is, the origin of coordinates is usually coincident with the core. But as we have discussed above, the core position depends on the frequency of observation and can move when new components are being ejected. Therefore, the positions of the components are actually referred to a location that is subject to change with epoch and frequency, and by an amount that it is unknown. However, there is a possible way to at least estimate these core motions. \emph{All} jet components are referred to the core position, and therefore any motion of the core through time or frequency should affect all of them by the \emph{same} amount. To search for these correlated motions we have considered the simplest case of constant rectilinear motion, which in fact is an excellent representation of the actual motion of the registering components, at least during the time covered by our 2001 observations. Core motions should appear as correlated deviations of the components from these simple linear fits. We should note, however, that linear motions of the core itself would obviously not be revealed by this procedure.

  Figure \ref{res} plots the residuals of linear fits in right ascension and declination for the registering components at the three frequencies. We have excluded the data for the last epoch in this analysis because of their poor quality. Figure \ref{res} clearly shows a pattern, far from the random behavior that would be expected in the case of changes in position caused by systematic errors in the model fitting, or by changes in the internal structure of the components. Accelerations, including bent trajectories, would also produce significant residuals, with a clear time evolution, as shown in Fig.~\ref{res}. However, it is highly unlikely that such deviations would be the same and occur simultaneously at two (three in the case of 15 GHz) components that are located in very different jet regions. (Components \emph{b} and \emph{c} move between approximately 1 and 3 mas from the core, while component \emph{f} moves between about 6.5 and 8 mas.) We therefore conclude that the residuals of Fig.~\ref{res} are representative of the core motion across epochs. To quantify this motion we have used the mean value of the residuals at each epoch, with an estimated error given by the standard deviation. The mean values of the errors are 15, 14, and 12 $\mu$as in right ascension, and 22, 21, and 30 $\mu$as in declination, at 15, 22 and 43 GHz, respectively.
  
  This procedure therefore provides an accuracy in the alignment of the images that is similar to what can be achieved by phase-reference VLBI astrometry \citep[e.g.,][]{Guirado:2000p13740} or SFPR \citep{Dodson:2009p14083}. Alignment of the images across frequencies is obtained through matching of the position of the optically thin registering components \emph{c} and \emph{f}.


\end{document}